\documentclass[twocolumn, switch]{article} % Method A for two-column formatting

\usepackage{preprint}

\usepackage{amsmath, amsthm, amssymb, amsfonts}

\usepackage[numbers,square]{natbib}
\usepackage[utf8]{inputenc}	% allow utf-8 input
\usepackage[T1]{fontenc}	% use 8-bit T1 fonts
\usepackage{xcolor}		% colors for hyperlinks
\usepackage[colorlinks = true,
            linkcolor = purple,
            urlcolor  = blue,
            citecolor = cyan,
            anchorcolor = black]{hyperref}	% Color links to references, figures, etc.
\usepackage{booktabs} 		% professional-quality tables
\usepackage{nicefrac}		% compact symbols for 1/2, etc.
\usepackage{microtype}		% microtypography
\usepackage{lineno}		% Line numbers
\usepackage{float}			% Allows for figures within multicol
\usepackage{siunitx}
\usepackage{framed,multirow}
\usepackage{newfloat}
\DeclareFloatingEnvironment[name={Supplementary Figure}]{suppfigure}
\usepackage{sidecap}
\sidecaptionvpos{figure}{c}

\usepackage{titlesec}
\titlespacing\section{0pt}{12pt plus 3pt minus 3pt}{1pt plus 1pt minus 1pt}
\titlespacing\subsection{0pt}{10pt plus 3pt minus 3pt}{1pt plus 1pt minus 1pt}
\titlespacing\subsubsection{0pt}{8pt plus 3pt minus 3pt}{1pt plus 1pt minus 1pt}

\usepackage{tikz,xcolor,hyperref}

\definecolor{lime}{HTML}{A6CE39}
\usepackage{wrapfig}
\newcommand{\InitialLetter}[1]{
\begin{wrapfigure}{l}{0.065\textwidth}
 \centering
 \vspace{-17pt}
 \includegraphics[width=0.09\textwidth]{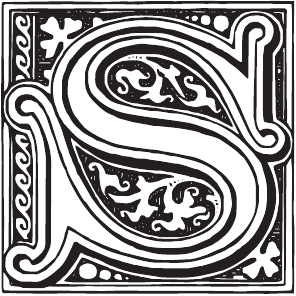}
  \vspace{-23pt}
\end{wrapfigure}
\noindent 
}
\title{PerSival: Neural-network-based visualisation for pervasive continuum-mechanical simulations in musculoskeletal biomechanics}

\usepackage{authblk}

\author[1\thanks{\tt{rosin@imsb.uni-stuttgart.de}}]{David Rosin}%\orcidA{}}
\author[2]{Johannes K\"assinger}
\author[3]{Xingyao Yu}
\author[4]{Okan Avci}
\author[5]{Christian Bleiler}
\author[6]{Oliver R\"ohrle}%\orcidB{}}

\affil[1,5,6]{Institute for Modelling and Simulation of Biomechanical Systems, University of Stuttgart}
\affil[2]{Institute for Parallel and Distributed Systems, University of Stuttgart}
\affil[3]{Visualization Research Center VISUS, University of Stuttgart}
\affil[4]{Biomechatronic Systems, Fraunhofer IPA, Stuttgart}
\affil[1,2,3,6]{Stuttgart Center for Simulation Science (SC SimTech), University of Stuttgart}

\begin{document}

\twocolumn[ % Method A for two-column formatting
  \begin{@twocolumnfalse} % Method A for two-column formatting
  
\maketitle

\begin{abstract}
This paper presents a novel neural network architecture for the purpose of pervasive visualisation of a 3D human upper limb musculoskeletal system model. Bringing simulation capabilities to resource-poor systems like mobile devices is of growing interest across many research fields, to widen applicability of methods and results. Until recently, this goal was thought to be out of reach for realistic continuum-mechanical simulations of musculoskeletal systems, due to prohibitive computational cost. Within this work we use a sparse grid surrogate to capture the surface deformation of the \textit{m.~biceps brachii} in order to train a deep learning model, used for real-time visualisation of the same muscle. Both these surrogate models take 5 muscle activation levels as input and output Cartesian coordinate vectors for each mesh node on the muscle's surface. Thus, the neural network architecture features a significantly lower input than output dimension. 5 muscle activation levels were sufficient to achieve an average error of 0.97 $\pm$ 0.16 mm, or 0.57 $\pm$ 0.10 \% for the 2809 mesh node positions of the biceps. The model achieved evaluation times of 9.88 ms per predicted deformation state on CPU only and 3.48 ms with GPU-support, leading to theoretical frame rates of 101 fps and 287 fps respectively. Deep learning surrogates thus provide a way to make continuum-mechanical simulations accessible for visual real-time applications.
\end{abstract}
%\keywords{First keyword \and Second keyword \and More} % (optional)
\vspace{0.35cm}

  \end{@twocolumnfalse} % Method A for two-column formatting
] % Method A for two-column formatting

%\begin{multicols}{2} % Method B for two-column formatting (doesn't play well with line numbers), comment out if using method A

%%%%%%%%%%%%%%%  Main text   %%%%%%%%%%%%%%%
% \linenumbers

\section{Introduction}
\InitialLetter{S}imulations enable the analysis of complex systems, even if they are inaccessible from an experimental point of view, be it for technical, conceptual, or ethical reasons. 
One such example is identifying tissue deformation due to forces exerted by or imposed on our musculoskeletal apparatus. 
Such data, however, is essential for tools intended to improve or contribute to computer aided surgery.
They rely on accurate and instantaneous predictions of a tissue's response to external manipulation.
\citep{sadeghi2020current,allard2007sofa}. 
Other examples are predicting joint movement due to muscle activation in the fields of rehabilitation, general physiotherapy or training \citep{ghannadi2017nonlinear}. 

If such applications appeal to biomechanical models at all, they typically appeal to simplifying lumped-parameter models such as those typically used in multi-body systems, e.g., \citep{hill1938heat, zajac1989muscle,meszaros2023effect}.
Even though they can provide results in real-time \citep{van2013real, ezati2019review}, they are not suitable for the above mentioned applications. 
Continuum-mechanical models that can provide information on tissue deformations are essential.
Depending on the level of detail of such models,~e.g.~\citep{bol2019investigating,klotz2020modelling}, considerable computing times might be needed - even if significant computational resources in form of supercomputers are available, cf.~\citep{bradley2018enabling, maier2019highly, Maier2021}.
Continuum-mechanical-based Finite Element (FE) simulations have the advantage over state-of-the-art Hill-type models that they can take the spatial heterogeneities and multi-physical phenomena into account.
Within such a framework, the governing equations of finite elasticity, realistic muscle geometries, and consistent initial and boundary conditions are discretised using the FE method \citep{rohrle2018skeletal,deBorst2018}. 
The solution of the resulting nonlinear system consists of the displacements of each nodal points of the FE mesh. 

Existing real-time and fully three-dimensional volumetric biomechanical simulations are proposed, for example, by \citeauthor{bro1996real} for surgical interventions involving the uterus, or by \citeauthor{mendizabal2020simulation}, a recent work on liver deformation. 
Besides individualised approaches to simulate three-dimensional tissue deformations in real time, different modelling environments have been developed over the last years, for example, Artisynth, which implements a hybrid approach between volumetric and lumped parameter model parts to focus computational resources \citep{lloyd2012artisynth}, or SOFA, which facilitates modularity with regards to more or less complex solvers and linear constitutive models \citep{allard2007sofa}. 
Both aim to optimise performance. 
Predicting and visualising muscle deformations in real-time based on continuum-mechanical musculoskeletal system modelling that also takes into account the muscular activation level, has not been achieved up till now.
Neither exist Virtual/Augmented Reality (VR/AR) environments that integrate such realistic, high-fidelity, biomechanical musculoskeletal system models into interactive systems.
Most VR/AR approaches build on a gaming engine or appeal to techniques typically developed for animations \citep{angles2019viper} neglecting the underlying biophysical principles. 

With this research, we aim to provide a proof of concept that on-body VR/AR visualisation of realistic muscle deformations resulting from a three-dimensional, volumetric biomechanical musculoskeletal system model are feasible in real-time ($\geq$30 frames per second) using a pervasive computing approach. 
Pervasive computing means (in this context) distributing computations and adopting model accuracy to achieve the required performance depending on the computational infrastructure, e.g.~on a mobile device or a server infrastructure. 
Our target application serves, aside from proof of concept, for more application-specific scenarios designed for medical staff or in therapeutic applications. 
Reliance on expensive, specialised hardware, stands in direct opposition to this. 
Hereby, surrogate modelling  aims to offer accurate prediction of the input-output-behaviour of a quantity of interest in a model, at a lower computational cost than evaluating the model itself. 
Beyond more traditional surrogate modelling methods, such as model order reduction techniques, Deep Learning (DL) approaches have been established in an effort to find abstract, real-time capable surrogate models for biomechanical simulations, as well as FE simulation in general \citep{tacc2023benchmarking,platzer2021finite,eggersmann2019model,holzapfel2021predictive}.
Deep Learning methods require large amounts of training data, which can, for example, be generated using offline simulations. 
Here, offline refers to running the necessary simulations completely independent of the target application and  on as large high-performance computing systems as needed. 

Within this work, we combine several surrogate modelling techniques.
Specifically, we adapt \citeauthor{valentin2018gradient}'s work to muscle surface deformation and use the resulting surrogate's output, which is based on a high-fidelity biomechanical FE musculoskeletal system model of the upper limb, as training data for a DL model (see also \citet{kneifl2023low}).
In other words, we simulate a high-fidelity model and use its output data to construct surrogate and DL models to achieve the maximum accuracy and evaluation speed possible with the respective and available hardware infrastructure. 
Further, we demonstrate that the performance is sufficient that the utilised high-fidelity model of the human upper limb can be visualised in real-time and on-body, using resource-constraint mobile devices such as VR/AR hardware.
In this context, we assume that data about a person's arm pose is provided in real-time and the neurological input to each muscle (one homogenised activation level per muscle) is known, e.g. as solution of an inverse problem that appeals to a sparse grid (SG) as a surrogate model.

\section{Methods}
Before going into the surrogate modelling process and its individual components, first the continuum-mechanical model of the upper arm will be described in some detail, as it is both the source of the data required for the proposed surrogate modelling/deep learning methods, as well as the starting point for all further considerations.

\subsection{Continuum-mechanical model}\label{FEsimulation}
The continuum-mechanical model of the upper limb consists of two main parts: a geometrical model discretized using the FE method and an appropriate constitutive formulation for the stress tensor describing the mechanical behaviour of the skeletal muscle tissue.
In brief, in continuum mechanics a body $\mathcal B$ is considered as a coherent manifold of material points (particles) $\mathcal{P}\in\mathcal{B}$. 
For later reference, we denote the body's surface as $\partial\mathcal B\subset\mathcal{B}$.
The material points $\mathcal{P}$ are parametrised by means of a position vector $\mathbf{X}$ in the undeformed reference configuration and the placement function $\chi (\mathcal P,t) =\chi (\mathbf{X},t) =  \mathbf{x}(\mathbf{X},t)$ establishes the mapping (point map) between the reference configuration and its current configuration with coordinates $\mathbf{x}$. 
Consequently, the displacement vector at time $t$ is defined as $\mathbf{u}(\mathbf{X},t) = \mathbf{x}(\mathbf{X},t)-\mathbf{X}$.
Further, the deformation gradient (line map) $\mathbf{F}$ is defined as
\begin{equation}
    \mathbf{F}(\mathbf{X},t) = \frac{\partial\chi(\mathbf{X},t)}{\partial \mathbf{X}} = \frac{\partial \mathbf{x}(\mathbf{X},t)}{\partial \mathbf{X}}\ \ .
\end{equation}
For later use, we further introduce the right Cauchy-Green deformation tensor $\mathbf{C} = \mathbf{F}^T\mathbf{F}$.

\begin{figure}
    \centering
    \includegraphics[width=0.35\textwidth]{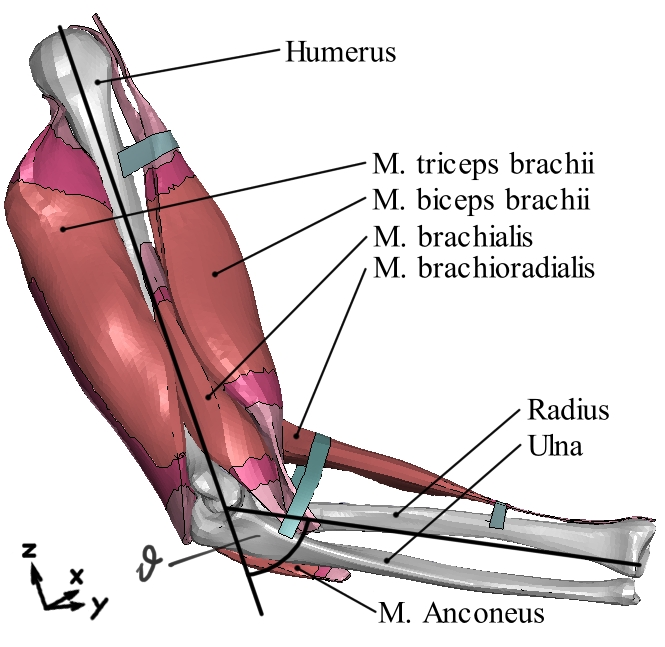}
    \caption{\textbf{Finite-element model.} Rendering of the finite-element upper-limb model, with the definition of the joint angle $\vartheta$. The colours help differentiate between pure muscle (brown-red), a muscle-tendon-mix (magenta), pure tendon (light red) and bands mimicing connective tissue (light blue).} 
    \label{fig:FE}
\end{figure}
The geometrical (FE) model of the upper limb contains three bones and five muscles (see Figure \ref{fig:FE}). The bones are modelled as rigid bodies and are namely the \textit{humerus}, \textit{radius}, and \textit{ulna}. 
\textit{Radius} and \textit{ulna} revolve around a hinge joint, connecting them to the lower head of the \textit{humerus}. The muscles actuating the joint are the \textit{m. biceps brachii} (biceps), \textit{m. brachialis} (brachialis), and \textit{m. brachioradialis} (brachioradialis) for flexion as well as the \textit{m. triceps brachii} (triceps) and \textit{m. anconeus} (anconeus) for extension. The distal end of the humerus is fixed and aligned with the z-axis.
The anatomical model of our upper limb (Figure \ref{fig:FE}) is based on the Visible Human dataset \citep{ackerman1998visible} and discretised using the FE method. The meshing was done with the pre-postprocessor ANSA. Special attention was given to ensuring sufficient accuracy of element density and element quality. The entire muscle-tendon-system consists of 638\,373 tetrahedral, quadratic Lagrange FE elements with 10 nodes each. The biceps by itself, which is in special consideration in this paper, is discretised with 68\,292 elements. For the mutual interaction between muscles and muscle with bone, a mortar surface-to-surface contact formulation is applied. The proximal tendons are fixed to the ulna and radius with tied node-to-surface contact formulation. 

The constitutive material description used within this work for the second Piola-Kirchhoff stress tensor $\mathbf{S}$ is based on a formulation using the right Cauchy Green tensor that is equivalent to the transversely isotropic muscle-tendon-description proposed by \citet{rohrle2017two}.
To describe the mechanical behaviour of the skeletal muscles constitutive behaviour, they use a phenomenological approach that is based on the principal invariants $I_1 = \rm tr[\mathbf{C}]$, $I_2 = \rm tr[cof[\mathbf{C}]]$, $I_3 = \det[\mathbf{C}]$ and the current (muscle) fibre stretch $\lambda_f =\sqrt{\mathbf{a}\cdot\mathbf{a}} = \sqrt{\rm tr[\mathbf{MC}]}$ with the structure tensor $\mathbf{M}=\mathbf{a}_0\otimes\mathbf{a}_0$, the referential fibre direction unit vector $\mathbf{a}_0$ and  ``$\otimes$'' denoting the dyadic product. 
The vector $\mathbf{a}_0$ relates to its current counterpart via $\mathbf{a}=\mathbf{F}\,\mathbf{a}_0$.
The stress $\mathbf{S}$ is additively split into an isotropic and an anisotropic term, i.e., $\mathbf{S}=\mathbf{S}_{iso}+\mathbf{S}_{aniso}$, and the anisotropic part is further split to take into account passive anisotropic tissue, e.g., tendon, and active contributions of skeletal muscle tissue, (cf. \citet{klotz2021physiology}):
\begin{flalign}\label{eq:S_split}
    \mathbf{S} = \mathbf{S}_{iso} + (\mathbf{S}_{passive} + \alpha\gamma_M\mathbf{S}_{active})(1-\gamma_{ST}),&&% + \OR{ p\mathbf{C}^{-1}},&&
\end{flalign}
where $\alpha\in [0,1]$ is the muscle activation, $\gamma_M, \gamma_{ST} \in [0,1]$ are factors to adjust the tissue compositions, e.g.~muscle %(\OR{$\gamma_M + \gamma_{ST}=1$ diese Bedingung ist mir nicht bekannt!}
with $\gamma_M, \gamma_{ST}>0$), tendon ($\gamma_M =0, 0<\gamma_{ST}\le 1$), and fat ($\gamma_M=\gamma_{ST}=0$). The isotropic part constitutes a Mooney-Rivlin-type material \citep{mooney1940theory,rivlin1948large}:
\begin{flalign}
    \mathbf{S}_{iso} = (B_1\mathbf{I}+B_2\mathbf{C}+B_3\mathbf{C}^{-1})+k\,(I_3^{1/2}-1)\,I_3^{1/2}\,\mathbf{C}^{-1},&&%2c_1(\mathbf{I} + \frac{1}{3}I_1\mathbf{C}^{-1}) + 2c_2(I_1\mathbf{I} - \mathbf{C} - \frac{2}{3}I_2\mathbf{C}) \ \ ,&&
\end{flalign}
where $k$ is the so-called bulk modulus \citep{crisfield1991nonlinear},
\begin{flalign}
    B_1 &= 2c_1I_3^{-1/3}+2c_2I_3^{-2/3}I_1,\\
    B_2 &= -2c_2I_3^{-2/3},\ \text{and}\\
    B_3 &= -2/3c_1I_3^{-1/3}I_1-4/3c_2I_3^{-2/3}I_2 .
\end{flalign}
with material parameters $c_1$ and $c_2$. 
As described in \cite{rohrle2017two}, the passive term $\mathbf{S}_{passive}$ of the anisotropic stress is chosen to be:
\begin{align}
    \mathbf{S}_{passive} =\ \begin{cases}
    \ \frac{c_3}{\lambda_f^2}\left(\lambda_f^{c_4} - 1\right)\mathbf{M} &\text{for}\ \lambda_f\geq1\ ,\\
    \ \mathbf{0} &\text{else}\ ,
    \end{cases}
\end{align}
where $c_3$ and $c_4$ are material parameters associated with the anisotropy of the material. The active part $\mathbf{S}_{aniso}$, is given by:
\begin{flalign}\label{eq:S_active}
    \mathbf{S}_{active} = \begin{cases}
    \frac{S_{max}}{\lambda_f^2}\exp{\left(-\left|\frac{\lambda_f/\lambda_f^{opt}-1}{\Delta W_{asc}}\right|^{\nu_{asc}}\right)}\,\mathbf{M} &\text{for}\ \lambda_f\leq\lambda_f^{opt}\\
    \frac{S_{max}}{\lambda_f^2}\exp{\left(-\left|\frac{\lambda_f/\lambda_f^{opt}-1}{\Delta W_{desc}}\right|^{\nu_{desc}}\right)}\,\mathbf{M} &\text{for}\ \lambda_f>\lambda_f^{opt}\ ,
    \end{cases}&&
\end{flalign}
where $S_{max}$ is the maximum active stress a muscle can generate at its optimal fibre length $\lambda_f^{opt}$. Furthermore $W_{asc}$ and $\nu_{asc}$ define the ascending part of the force-length-relation of the muscle, while $W_{desc}$ and $\nu_{desc}$ are the corresponding counterparts for the descending part. 
The material parameters are summarised in Table \ref{tab:params} in the Appendix.

The constitutive material model defined in Eqs~\eqref{eq:S_split}-\eqref{eq:S_active} has been implemented as a user-material in the commercial FE software LS-DYNA\footnote{https://lsdyna.ansys.com}.
An implicit solver is used to solve the appropriate weak form of the linear momentum balance (the governing equation)
\begin{equation}
	\begin{aligned}
		 & \textsf{div}\,\mathbf{\sigma}+\rho\,\mathbf{g} = \rho\,\mathbf{\ddot{x}}, %\quad \textrm{with} & \rho = \frac{1}{\textsf{det}\,\mathbf{F}}\,\rho_0,
		 \label{eqn:balancelinearmomentum}
	\end{aligned}
\end{equation}
in spatial configuration.
The stress tensor $\mathbf{S}$ from Eq.~\eqref{eq:S_split} is connected to the Cauchy stress tensor in Eq.~\eqref{eqn:balancelinearmomentum} via $\mathbf{\sigma} = \det[\mathbf{F}]^{-1}\ \mathbf{F}\, \mathbf{S}\, \mathbf{F}^{T}$.
Further, $\rho$ is the material density in the current configuration and $\mathbf{g}$ denotes acceleration
due to gravity.
We expect sufficiently slow movements, $\mathbf{\ddot{x}}\approx\mathbf{0}$, so that the problem becomes quasi-static.
To ensure nearly incompressible deformations, $k$ is chosen sufficiently large.\\
The continuum-mechanical model described in this section constitutes a function $\pmb \alpha \rightarrow \mathbf{x}(\mathcal P),\ \forall\ \mathcal P \in \mathcal B$ and $\forall\ \alpha_i \in [0,1],\ i=1,..,5$, where $\alpha_i$ refers to the activation levels of the individual muscles in the model, while $\pmb\alpha$ is a vector containing all five activation values. For the subsequent surrogate modelling we will - without loss of generality - focus on the biceps surface specifically.

\subsection{Sparse grid surrogate}\label{SGsurrogate}
With respect to the elbow joint angle the upper limb model represents an overdetermined system. Consequently an optimisation is needed to solve the inverse problem of finding the muscle activation levels necessary to reach a given pose.
The original FE model cannot be used to handle this process in a reasonable time span, not even on high-performance hardware. 
Tests on a cluster with 64 CPUs, with 3 GHz each, and 1 TB of RAM showed evaluation times of up to 4 minutes per deformation state. Instead, FE model evaluations were restricted to 1053 states, uniformly distributed across the 5D parameter space of all activation levels.
The number of states and their distribution was chosen to fit the support for a uniform SG. 
We use this surrogate to interpolate the known activation states, in order to evaluate the displacement for any activation $\alpha_i \in [0,1]$. 
The SG only interpolates in the activation-domain not in space. Strictly speaking we use an ensemble of SGs, one per node in the FE mesh. 
For the purposes of this paper, we simply refer to this ensemble as a whole as the SG surrogate, which constitutes a function $\alpha \rightarrow \mathbf{x}(P_{mesh}),\ \forall\ P_{mesh} \in \partial\mathcal{B}_{biceps}$, where $P_{mesh}$ refers to mesh node, rather than material points $\mathcal P$. 
The size of the support of a SG is characterised by its level $l$. A higher level results in a more fine-grained grid, allowing for more accurate interpolation. 
Choosing a level is thus a trade-off between computational cost and accuracy. 
Based on the previous findings by \citeauthor{valentin2018gradient} a regular SG of level $l=2$ was deemed dense enough to achieve sufficient approximation accuracy. Each activation level of one of the five muscles in the model constitutes one dimension of the SG. Spanning this parameter space with a SG of level 2 requires the aforementioned 1053 unique activation combinations.
The interpolation between the support points is based on cardinal B-splines of degree $p = 3$.
A B-spline $b^p : \mathbb{R}\to\mathbb{R}$ is defined recursively by:
\begin{align}
    b^0(x)= 1_{[0,1)}(x),\ \ b^p(x) = \int_0^1b^{p-1}(x-y)dy\ \ .
\end{align}
The hierarchisation of these splines is given as the affine transformation
\begin{align}
    \varphi^p_{l,i}(x) = b^p\left(\frac{x}{h_l} + \frac{p+1}{2} - i\right),\ \ h_l = 2^{-l}\ \ .
\end{align}
For SGs of dimension $d$, multivariate, hierarchical B-splines are obtained by computing %using 
a tensor product
\begin{flalign}
    \varphi^{\mathbf{p}}_{\mathbf{l},\mathbf{i}}(\mathbf{x}) = \prod^d_{t=1}\varphi^{p_t}_{l_t,i_t}(x_t),\ \ \mathbf{l}\in \mathbb{N}^d,\ \mathbf{i}\in I_{\mathbf{l}} = I_{l_1}\times,...,\times I_{l_d}\ \ ,&&
\end{flalign}
with corresponding grid-points
\begin{align}
    \mathbf{x}_{\mathbf{l},\mathbf{i}} = (x_{l_t,i_t})_{t=1,..,d},\ \ x_{l_t,i_t}=i_t\cdot h_{l_t}\ \ .   
\end{align}
In the multivariate case $\mathbf{i}$ and $\mathbf{l}$ encompass all indices and levels associated with a given quantity, as both of those are defined per input dimension and can vary between them.
\citeauthor{valentin2016hierarchical} showed that  for odd $p$ the nodal subspace $V_l^p$ of a regular SG of level $l$ and dimension $d$ can be described by the direct sum:
\begin{align}
    V_{\mathbf{l}}^{\mathbf{p}} = \underset{\parallel \mathbf{l}\parallel_1\leq n+d-1}{\oplus}\text{span}\{\varphi^{\mathbf{p}}_{\mathbf{l},\mathbf{i}}|\mathbf{i}\in I_{\mathbf{l}}\}
    %V_l^p = \oplus_{\parallel \overrightarrow{l}\parallel_1\leq n+d-1}\text{span}\{\varphi\}
\end{align}
The contribution of each subspace to the approximation $\hat f$ of a given function $f:[0,1]\to\mathbb{R}$ is controlled by weighting-coefficients, the so-called hierarchical surpluses $\alpha_{l',i'}$, and defined by
\begin{align}
    \hat f (\mathbf{x}) = \sum_{l'=0}^l\sum_{i'\in I_{l'}}\alpha_{l',i'}\varphi_{l',i'}(\mathbf{x}) \ ,
\end{align}
with
\begin{align}
    \hat f(x_{l,i}) = f(x_{l,i})\ \ \forall i \in \left[0,2^l\right] \ .
\end{align}

\subsection{Neural network surrogate}\label{NNsurrogate}
With sparse-grid-surrogate evaluation times of up to 100 ms for the visualisation of the biceps' surface, the SG surrogate is significantly faster than the original FE simulation but not fast enough for real-time visualisation. Thus results from SG interpolation are used as training data for a real-time-capable deep learning surrogate. 
To predict the activation- and stretch-induced deformation of the biceps' surface in real-time, a densely connected feed-forward neural network (NN) is employed. 
Note, the NN surrogate is not supposed to be a one-to-one replacement for the SG. 
As the relation between joint angle and muscle activation input might not be bijective, i.e.~ multiple joint angles might relate to the same muscle activation levels, we additionally desire "energy efficiency of the musculoskeletal system", i.e.~penalising  sets of activation levels with high values. 

Within this work, we achieve this by defining the following cost function for the optimisation:
\begin{equation}
    \pmb\alpha^{\mbox{\scriptsize k}}_{\mbox{\scriptsize opt}} = \underset{\pmb\alpha^{\mbox{\tiny k}}\in[0,1]^5}{\text{argmin}}\,\left(\left|{\varTheta^{\mbox{\scriptsize k}}}-\vartheta^{\mbox{\scriptsize k}}(\pmb\alpha^{\mbox{\scriptsize k}})\right| + 0.05 \cdot \sum_{i=1}^5\alpha^{\mbox{\scriptsize k}}_i\right)\ \ ,
    \label{alphaopt}
\end{equation}
where $\varTheta^{\mbox{\scriptsize k}}$ is the target elbow angle and $\vartheta(\pmb\alpha^{\mbox{\scriptsize k}})$ is the in silico determined elbow angle due to activation parameters $\pmb\alpha^{\mbox{\scriptsize k}}$. 
The scaling of the sum of the activation levels has been chosen small, i.e.~$0.05$, keeping the focus of the minimisation on target angle $\varTheta^{\mbox{\scriptsize k}}$.  

To speed up the activation-joint-angle-relationship calculations during the online phase, we evaluate  equation (\ref{alphaopt}) for a set of admissible activation parameters ($\SI{10}{\degree}\le\varTheta^{\mbox{\scriptsize k}}\le\SI{150}{\degree}$) and subsequently use the $\pmb\alpha^{\mbox{\scriptsize k}}_{\mbox{\scriptsize opt}}$ within a SG surrogate for the deformation of each mesh node on the muscles' surface. % x-x
As we consider quasi-static conditions, we omit superscript $k$ for the time instance from now on.
The deformation is then used to train a NN surrogate, i.e.~it is trained to approximate a function $\pmb\alpha_{\text{opt}} \rightarrow \mathbf{u}(P_{mesh}), \forall\ P_{mesh}\in \partial \mathcal B_{biceps}$.
This is obtained by evaluating the SG surrogate for all the mesh-nodes $P_{mesh}$ at $\pmb\alpha_{\mbox{\scriptsize opt}}$, and subtracting the known reference coordinates of the given node $\mathbf{X}(P_{mesh})$ from the result.
All these evaluations and the training of the network can be done during the offline phase. 
The reference configuration is added to the bias of the last layer in the NN post-training, to convert from the displacement back to the Cartesian coordinates of the mesh-nodes. 
With this the NN surrogate provides the desired function $\pmb\alpha_{\text{opt}} \rightarrow \mathbf{x}(P_{mesh}), \forall\ P_{mesh}\in \partial \mathcal B_{biceps}$.

The NN was implemented using the open-source API Keras for TensorFlow in Python \citep{tensorflow2015-whitepaper,chollet2015keras}.\\

\subsubsection{Neural Network Architecture}
\begin{figure}
    \centering
    \includegraphics[width=0.48\textwidth]{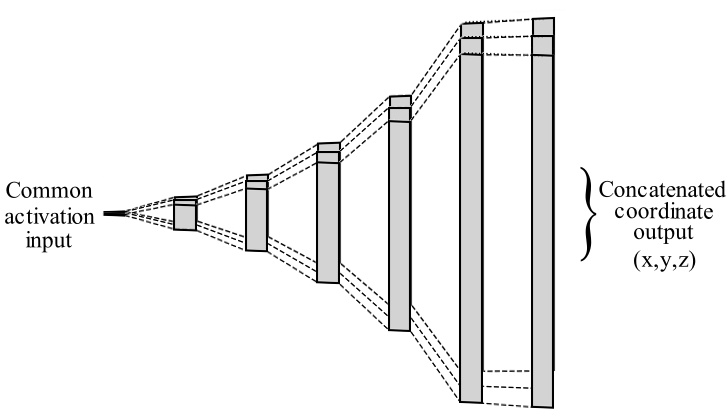}
    \caption{\textbf{Combined model.} Combination of the three NNs for the x-, y- and z-coordinate, by defining a common input for all three networks and a concatenated output.}
    \label{fig:NN_comb}
\end{figure}
The NN architecture needs to bridge the gap between the number of inputs $n_{in} = 5$, i.e.~the number of muscle activation parameters of the FE model, and the number of outputs $n_{out} = 2809$. 
The NN consists entirely of densely connected layers. 
While there are three Cartesian coordinates to consider for each node, the coordinate directions are trained separately, each using an identical architecture. 
The reasons for this will be discussed in more detail in the next section. 
The width $w$ of each layer, i.e.~the number of neurons per layer, increases exponentially from input to output of the NN. 
The parameters of the exponential function are the number of hidden layers $n_{hidden}$, $n_{inputs}$, $n_{outputs}$, and $n_{mid}$, which specifies the width to be reached half-way between the input- and output-layer: 
\begin{align}
    w(n) &= a \cdot e^{b \cdot n} + c\ ,\ \text{with}\\
    a &= \frac{n_{mid}^2 - n_{inputs}^2}{n_{outputs} - 2\cdot n_{mid} - n_{inputs}}\ ,\\
    b &= 2\cdot n_{hidden}\cdot\log(\frac{n_{outputs} + n_{inputs}}{n_{mid} + n_{inputs}})\ ,\\
    c &= n_{inputs} - a\ \ .
\label{eq:shape}
\end{align}
Figure \ref{fig:NN_comb} shows a visual representation of this NN architecture.
The best training results were reached with the number of hidden layers $n_{hidden}  = 5$ and the width of the mid-layer of the network $n_{mid} = 700$. 
These parameters were chosen as small as possible, while maintaining sufficient accuracy, i.e.~sufficiently low mean squared error (MSE), of the predicted coordinate output, by comparing NNs with varying values for these parameters. 

\subsubsection{Dimension-decoupled training}
Three individual networks were trained, one for each coordinate dimension. 
Post-training, these individual models were combined into a single NN via network-graph redefinition, using the Keras functional API \ref{fig:NN_comb}). 
Each of the three networks was trained using five-fold cross-validation \citep{ojala2010permutation}, i.e.~the training data was shuffled, and divided into five partitions, one of which was used for validation and four for training for a certain number of epochs defined. 
After each ``fold'', which partition is used for validation and thus not shown to the model during training changes. 
The networks were trained for 400 epochs on each fold, resulting in an overall training duration of 2000 epochs on the MSE, with the mean absolute error (MAE) as an additional metric. 
Here ``mean'' refers to a mean value across all surface nodes in a given configuration. 
Additionally Keras' ``callbacks''-functionality was used to select the version of the NN that performed best, i.e.~has the lowest loss across all epochs. 
This was done to account for fluctuations in the in-training performance from epoch to epoch, meaning the last iteration of the NN was not necessarily the best performing one. 
Keras' functional API allows for the redefinition of network-graphs on a layer-by-layer basis. 
This was used to define a common input and output for the three networks, without needing to manipulate their tensors manually. 
The only manual weight-editing is the aforementioned addition of the reference configuration to the last layer's bias, converting the NN's output from displacement to Cartesian coordinates.

\section{Results}

\subsection{Performance results}
\begin{figure}
    \centering
    \includegraphics[width=0.48\textwidth]{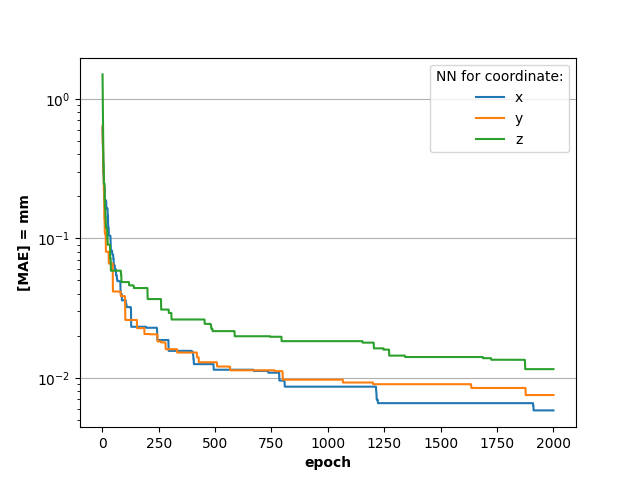}
    \caption{\textbf{Training history.} In-training performance of the three coordinate-networks shown as the MAE in mm over the number of epochs trained.}
    \label{fig:MAE_over_epoch}
\end{figure}
Figure \ref{fig:MAE_over_epoch} shows the performance of the current lowest loss model over the course of 2000 epochs of training. 
For reference, without the use of callbacks the model loss fluctuated  between $1 \cdot 10^{-1}$ and $1 \cdot 10^{-2}$ mm of MAE. 
The initial drop in loss over the first 50 epochs was not subject to these fluctuations.
\begin{figure}
    \centering
    \includegraphics[width=0.48\textwidth]{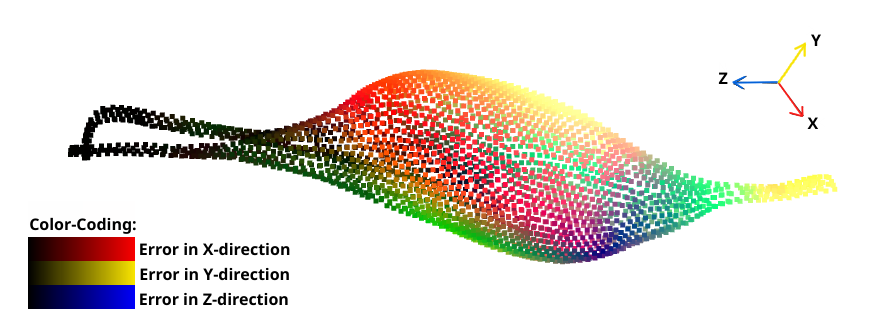}
    \caption{\textbf{Error ``hot-spots''.} Qualitative error analysis after 100 epochs of training, with colour-coding for x (red), y (yellow), z (blue). Yellow regions of comparatively high error appear in multiple configurations.}
    \label{fig:quali_err}
\end{figure}
In the intermediate results from the first 100 epochs of training the accuracy of the predicted node coordinates differed significantly across the biceps' surface. 
More specifically, while visualising a cycle of arm flexion and extension, consistent areas of high absolute error.
This did not correspond to a higher MAE than for other surface configurations. 
This mostly concerned configurations for which all flexors were at least partially %, but not fully 
activated. 
An example of a configuration exhibiting these hot-spots can be seen in Figure \ref{fig:quali_err}, where the higher the MAE, the lighter the colour of the area. 
These regions, in order of decreasing MAE, are namely the lower tendon, the medial side of the biceps' belly, as well as the more lateral, dorsal area of the muscle belly. 
The error distribution across the muscle equalised over the rest of training and hot-spots were no longer present by the end of it. 
In-training accuracy after 2000 epochs showed a maximum absolute error (MaxAE) of 0.37 mm, which corresponds to 0.38\%, with a mean error of 0.014 mm.
All three of these values give an average over all evaluated states, with mean and maximum taken for each.\\
Out-of-training performance was tested by iteratively generating random activation vectors for which both the NN and the SG were evaluated. 
Out-of-training performance refers to a NN generalisation capabilities, i.e.~generating output for inputs that were not included in the training data. 
This still holds true in our case, but does not include any extrapolation, since our input parameter space is strictly bounded. 
New random activation vectors were tested until the highest recorded MaxAE of the NN w.r.t. the SG no longer increased for 300 consecutive iterations, i.e.~until a worst-case accuracy was found. 
The results were a maximum MaxAE of 8.82 mm and maximum relative error (MaxRE) of 10.12\%. 
Here, the MaxRE is not the converted MaxAE, but rather a separately recorded maximum value. 
The highest MAE across all configurations was 1.69 mm, the highest mean relative error (MRE) 1.23\%.\\
Random sampling will include activation vectors, which will not be encountered in the final application, as they are not an optimal choice for any given joint angle. 
Instead iteratively testing for random joint angles, for which the optimal activations are computed, yields a MaxAE of 3.75 mm and a MaxRE of 5.69\%, a MAE of 0.97 mm and a MRE of 0.57\%. See Table \ref{tab:out_of_training} for a summary of these results. 
\begin{table}[h]
    \centering
    \caption{\textbf{Out-of-training accuracy.} Summary of the out-of-training accuracy of the NN tested on random activations and optimal activations for random joint angles. Relative erros were determined independently from the absolute values.}
    \label{tab:out_of_training}
    \begin{tabular}{|c|c|c|c|c|}
        \hline
         & \multicolumn{2}{c|}{\textbf{maximum error}} & \multicolumn{2}{c|}{\textbf{mean error}}\\
        % & abs. & rel. & abs. & rel.\\
        \textbf{tested for:} & mm & \% & mm & \%\\
        \hline
        % new values for this line: 11.96 mm 13.28 % 2.15 mm 1.59 %
        \textbf{random} & \multirow{2}{1cm}{11.96} & \multirow{2}{1cm}{13.28} & \multirow{2}{1cm}{2.15} & \multirow{2}{1cm}{1.59} \\
        \textbf{activation} & & & & \\
        \hline
        \textbf{random} & \multirow{2}{1cm}{ 3.75} & \multirow{2}{1cm}{ 5.69} & \multirow{2}{1cm}{ 0.97} & \multirow{2}{1cm}{ 0.57} \\
        \textbf{angle} & & & & \\
        \hline
    \end{tabular}
\end{table}

The model was also tested for evaluation speed. 
This was done for sequential execution of the three separate NNs for the x-, y- and z-coordinate, the combined model as shown in Figure \ref{fig:NN_comb}, and a TensorFlow-Lite-version of the combined model. 
The models were tested on CPU only (Intel i5 with 6 cores with 3 GHz) and with GPU support (Nvidia Geforce GTX 1650 with 4GB VRAM).
\begin{table}
    \centering
    \caption{\textbf{Average evaluation times.} Comparison of evaluation times of different versions of the NN model, averaged over 1000 evaluations.}
    \begin{tabular}{|c|c|}
        \hline
         \textbf{Model version} & \textbf{Evaluation time in ms} \\
         \hline
         three separate models & 13.65\\
         $\to$ with GPU & 4.05\\
         \hline
         combined model & 13.34\\
         $\to$ with GPU & 3.48\\
         \hline
         combined model Lite & 9.88\\
         \hline
    \end{tabular}
    \label{tab:eval_times}
\end{table}
Table \ref{tab:eval_times} shows the evaluation times averaged over 1000 evaluations are approximately the same for the sequential and the combined model when executed on the CPU, with a theoretical frame rate of about 75 frames per second (fps). 
For execution on the GPU the times differs, with the combined model being 0.57 ms faster than the sequentially executed models.
The theoretically achievable frame rates are 246.91 fps and 287.36 fps respectively. 
The results for the TensorFlow Lite model are identical for GPU and CPU, with an evaluation time of 9.88 ms, i.e.~101.21 fps.

\subsection{Error Propagation}
Since the NN surrogates are trained on data from the SG surrogate, not the FE simulation, i.e.~approximating based on an approximation, error propagation should be considered. 
The worst case scenario would be that the NN amplifies the approximation error of the SG. 
To investigate this the FE model was run for 25 activation combinations drawn randomly from a uniform distribution. 
Both the SG and the NN surrogate were evaluated for the same activation vectors and the results of the subsequent comparison of both to the FE simulation are summarised in Table \ref{tab:SG_val}.
\begin{table}[h]
    \centering
    \caption{\textbf{Error propagation test.} Summary of the accuracy of the SG and the NN tested on 25 random activations evaluated using the FE model. The NN was also tested on the 1053 support-points of the SG, since those are taken from the FE-model as well.}
    \label{tab:SG_val}
    \begin{tabular}{|c|c|c|c|c|}
        \hline
         & \multicolumn{2}{c|}{\textbf{maximum error}} & \multicolumn{2}{c|}{\textbf{mean error}}\\
        % & abs. & rel. & abs. & rel.\\
        \textbf{Surrogate:} & mm & \% & mm & \%\\
        \hline
        % new values for this line: 11.96 mm 13.28 % 2.15 mm 1.59 %
        \textbf{SG} & 17.79 & 10.14 & 2.86 & 1.52 \\
        \hline
        \textbf{NN} & 18.28 & 9.38 & 2.84 & 1.45 \\
        \hline
    \end{tabular}
\end{table}
The error of the NN with respect to the original FE simulation closely matches that of the SG, both with significant inaccuracies. 
Put differently, the SG appears to be the primary contributor to the overall approximation error of the surrogate system, while the NN does not notably add to it (see also \ref{tab:out_of_training}).

\section{Discussion}
Results derived using short training times revealed error hot-spots, as shown in Figure \ref{fig:quali_err}, which were found to be consistent in most surface configurations. 
These regions of comparatively high error on the biceps' belly coincide with contact areas with other flexors, like the \textit{M. brachioradialis} (BR). 
A possible explanation for the apparent increased learning complexity in these regions is contact to the BR deflecting surface nodes from a more common trajectory. 
Trajectories of nodes in these regions are likely more unique compared to the majority of nodes. 
Deflection of the biceps by the BR is likely aggravated by the lack of a fat-skin-layer surrounding the musculature, causing the BR to contract less in line with the humerus and biceps, than it realistically would. 
Implementation of a fat-skin-layer is already planned for future versions of this model. 
The tip of the lower tendon of the biceps is also an area of comparatively high error, which is caused by its attachment to the radius bone. 
Movement of these nodes is determined by the rotation of the lower arm around the elbow joint, which again results in node trajectories unlike the majority throughout the muscle. 
Deep learning models, especially classifiers, which sort data into categories, are known to have difficulties learning rare cases in their training data. 
Data augmentation techniques can be used to make rare occurrences more prevalent in the data, increasing their influence during training \citep{lim2018doping}. 
A similar effect could be achieved for our regression model, e.g.~by subdividing the NN model into parts specialising in these regions and nodes.
However, since the hot-spots vanish given sufficient training time, data augementation will not be necessary in this case. It is interesting to see muscle-to-muscle contact have this effect nonetheless. 

\subsection{Out-of-Training Performance}
Out-of-training accuracy was within expectations. 
While the NN can not provide the level of accuracy of the original simulation or the in-training accuracy, even the worst case accuracy shown in Table \ref{tab:out_of_training}, with 5.69\% MaxRE and 3.75 mm MaxAE, are not of concern for visualisation-purposes. 
Especially considering the MAE and MRE are an order of magnitude below the worst case performance, with 0.97 mm and 0.57\%. 
The final application envisioned by this project will include additional, external computational resources, as well as a number of models with different complexity and accuracy. 
Which model to use, and how heavily computations are offloaded, will be decided by the application at run-time, to provide the best possible performance depending on the available hardware. 
The additional results regarding error propagation shown in Table \ref{tab:SG_val} show inaccuracies of the SG to dominate the overall error. 
These inaccuracies can not be disregarded as fringe effects, as they occur at mid level activations. 
The current SG is uniform in all five activation dimensions. 
This can be improved upon by building the SG support adaptively, which involves evaluating SG accuracy while building its support by running the FE-model, iteratively adding support points as and where needed \citep{valentin2019b}. 
Which points are most likely to improve accuracy the most can be estimated, as for example demonstrated by \citeauthor{khakhutskyy2016spatially}. 
An adaptive SG will be implemented in next version of this setup.\\
There is an overarching point to be made about error propagation with coupled surrogate models, as we use them here. 
Deep learning models are black-boxes with no formal way for predicting their accuracy, and are highly dependent on the quality of their training data \citep{miotto2018deep}. 
Continuum mechanics relies on homogenisation of the micro-scale, while the FE method  introduces a discretisation error. 
Care needs to be taken, when chaining models and methods in this manner, with a data driven model as the endpoint. 
This is however mitigated by the fact that the strength and weaknesses of both the FE method, continuum mechanics, and SG are well and formally established \citep{bonet1997nonlinear,wriggers2013nichtlineare}.\\
The measured evaluation times (see Table \ref{tab:eval_times}) shows that all tested versions of the model are real-time-capable. 
A conversion to an optimised format, such as TensorFlow Lite, is preferable, if no GPU is available. 
TensorFlow Lite is optimised for ARM CPUs, to better fit mobile devices, where they are commonly used. 
Thus, the results listed here, likely do not show the peak performance of the TensorFlow Lite version of the model, since the Intel CPU we tested on is of a different architecture.\\
It is already known from preliminary tests that running the NN model will not be the biggest contributor to the overall computational load of the target application. 
This will rather be the motion tracking needed to estimate the pose of a person the model is supposed to be overlayed on in VR/AR, as well as the visualisation of the model. 
Fortunately all versions of the model tested here lie well below the guideline for real-time applications of 30 ms evaluation time, leaving us with computational resources to spare at the current accuracy and speed. 

\subsection{Architecture Design process}
Since a significant part of the results of this work is the specialised NN-architecture, we would like to give some insight into its construction.\\
Fully connected layers were used for this model, as there was nothing about the data indicating an immediate benefit from using a more specialised layer-type. 
It also offered the most flexibility in shaping the architecture overall. 
To connect the five input parameters to the 2809 outputs, three rudimentary options came to mind:
\begin{enumerate}
    \item changing the layer-width as late as possible, resulting in a T-shaped NN
    \item changing the layer-width as early as possible, resulting in a rectangular NN
    \item changing the layer-width linearly as a function of the distance to the input, resulting in a triangular NN
\end{enumerate}
To be able to compare these three architectures, prototypes were generated with an arbitrarily chosen 10 hidden layers.% and trained for the same number of epochs.
\begin{table}[h]
    \centering
    \caption{\textbf{Prototyping.} Comparison of three different prototype-architectures for learning the X-coordinates of the biceps' mesh, based on in-training MAE after 10 epochs, and number of trainable parameters.}
    \begin{tabular}{|c|c|c|c|}
        \hline
         & \textbf{T-shaped} & \textbf{rectangular} & \textbf{triangular}  \\
         \hline
        \textbf{\# of Parameters} & 17.154 & 71.056.494 & 26.103.046 \\
        \hline
        \textbf{MAE in mm} & 430 & 2.4 & 2.2\\
        \hline
    \end{tabular}
    \label{tab:prototypes}
\end{table}
As this was only the initial prototyping stage, short training times of only 10 epochs were deemed sufficient to check if a particular network shape would lend itself to the data. 
An example of such a test can be seen in Table \ref{tab:prototypes}. 
In this comparison the T-shaped architecture performs significantly worse than its competitors. 
The rectangular and triangular architectures are on par. 
However the triangular architecture reaches this level of accuracy with less than half the number of trainable parameters of the rectangular one. 
The exponential architecture introduced in the Methods section was then developed in an attempt to further decrease the number oftrainable parameters, by slimming its mid-section, without loss in accuracy. 
The number of hidden layer was then iteratively decreased down to five, at which point further reduction caused a notable decline in training accuracy. 
The same was done for the width of the mid-layer of the model (see \ref{eq:shape}). \\
These initial steps were taken while training on the nodal coordinates, not the deformation. 
However, first visualisations using this method, revealed the biceps to slightly drift sideways during contraction. 
Switching to training on the deformation instead successfully removed any drift from the predicted nodal positions. 
Further improvements were then simply achieved with longer training times.\\
All these steps could be done by considering only one of the three output dimensions. 
When trying to expand the architecture to 3D, with the same number of hidden layers and shape, training performance was significantly worse. 
To circumvent this problem, without increasing the size of the NN, three individual networks were trained, one for each coordinate dimension, and combined into a single NN via network-graph redefinition post-training.

\section{Conclusion and outlook}
The aim of this work was to find a work flow that is suitable for visualising data from continuum-mechanical forward simulations in real-time, to further the development of pervasive simulation and open up continuum mechanics to a wider variety of applications. 
The presented proof-of-concept model is able to predict the surface configuration of the biceps, based on a vector of activation levels optimal for reaching a certain target elbow angle, within an average error of 0.97 mm at a frame rate of up to 287.35 fps on non-high-end hardware. 
This model thus fulfills our set criteria. 
Also note that even taking the aim of pervasiveness aside, real-time prediction of the muscle's surface configuration on a non-high-end PC, is still an exciting result, considering the original FE simulation took minutes per configuration on 64 CPUs à 3GHz, with 1 TB of RAM. 
The use of a SG surrogate to generate the training data for the NN, has proved  efficient for acquiring arbitrary amounts of training data within a feasible time frame. 
Accuracy of the SG still needs to be improved by optimising its support. 
Next steps are the expansion of this workflow to the full FE model, including five muscles and three bones, contributing to a complete simulation of the upper arm.\\
These models will be implemented and extensively tested on resource-constraint devices, e.g.~a distributed AR application on the Microsoft HoloLens 2. 
To achieve this, firstly a second model trained on the relationship between the motion tracking data (elbow angle, angular velocity and acceleration) and the optimal activation vectors will be introduced; secondly, once the output visualisation is overlayed onto a person, it will be supplemented with additional biomechanical information, e.g.~graphs tracking the activation levels. 
A scenario these two steps will be relevant for is for example coupling the model to a motion guidance system for physiotherapy or exercise.\\
For instance, in the physiotherapy setting, a doctor asks their patient to complete an arm movement following a guided path, shown to them in VR/AR. 
This model could help the patient to appreciate how their motions result from muscle contraction and aid the physiotherapist in explaining possible issues and solutions. 
Ideally, this could allow the patient to share in their therapist's intuition as to what needs to be done. 
The guidance could help provide feedback to the patient on what exactly the motion is, they should try to achieve \citep{han2016ar,yu2020perspective}, even continue to do so, when training unsupervised at home \citep{tang2015physio}. 
Further development of this model beyond the technical aspects will include incorporating it into new on-body interaction techniques, to not only further improve performance on a wider range of hardware, but also provide ease of use to the end-user.\\
Finally, we would like to note that the methods presented in this paper are neither limited to the muscle constitutive laws chosen in Section~\ref{FEsimulation} nor to FE results in general. 
The presented sparse-grid-neural-network workflow is equally applicable to more complex, microstructural-based muscle constitutive laws \citep{Bleiler2019} or activation mechanisms \citep{klotz2021physiology}, as well as to other biological or man-made materials. 
Likewise, the method can be applied to any kind of mesh-based data, such as results coming from the finite-difference-method.

\appendix
\section{Constitutive parameters}
\setcounter{table}{0}
\begin{table}[h]
    \centering
    \caption{\textbf{Model parameters.} Constitutive parameters for the muscle-tendon-complex of the biceps.}
    \begin{tabular}{|c|c|c|}
        \hline
        \textbf{Parameter} & \textbf{Value} & \textbf{Source}\\
        \hline
         $c^M_1$ & $3.56\cdot 10^{-2}$ MPa & \multirow{2}{2cm}{\citet{hawkins1994comprehensive}}\\
         %\cline{1-2}
         $c^M_2$ & $3.86\cdot 10^{-3}$ MPa & \\
         \hline
         $c^M_3$ & $3.57\cdot 10^{-8}$ MPa & \multirow{2}{2cm}{\citet{zheng1999objective}}\\
         %\cline{1-2}
         $c^M_4$ & 42.6 & \\
         \hline
         $c^T_1$ & 2.31 MPa & \multirow{4}{2cm}{\citet{weiss2001computational}}\\
         %\cline{1-2}
         $c^T_2$ & $1.15\cdot 10^{-6}$ MPa & \\
         %\hline
         $c^T_3$ & 7.99 MPa & \\%\multirow{2}{2cm}{Weiss and Gardiner}\\
         %\cline{1-2}
         $c^T_4$ & 16.6 & \\
         \hline
         $\Delta W_{asc}$ & 0.25 & \multirow{6}{2cm}{adapted from \citet{gunther2007high}}\\
         %\cline{1-2}
         $\Delta W_{desc}$ & 0.15 & \\
         %\cline{1-2}
         $\nu_{asc}$ & 3.00 & \\
         %\cline{1-2}
         $\nu_{desc}$ & 4.00 & \\
         %\cline{1-2}
         $\lambda_opt^f$ & 1.35 & \\
         %\cline{1-2}
         $\sigma_{max}$ & 0.30 MPa & \\
         \hline
    \end{tabular}
    \label{tab:params}
\end{table}

\section*{Acknowledgments}
\noindent
This research was partially funded by the Deutsche Forschungsgemeinschaft (DFG, German Research Foundation) under Germany's Excellence Strategy - EXC 2075 – 390740016, the Stuttgart Center for Simulation Science (SimTech)\\
\noindent
Fraunhofer's Internal Program under Grant No. MAVO 828 424  and the Bundesministerium für Bildung und Forschung (BMBF, German Federal Ministry of Education and Research) under Grant No. 01EC1907B (“3DFoot”).\\
\noindent
LaTeX-template for this preprint available at: https://github.com/brenhinkeller/preprint-template.tex under license CC by 4.0.

\section*{Data availability}
\noindent
Sparse grid data, training data, and code for the models presented here will be made available on https://darus.uni-stuttgart.de upon peer-reviewed publication of this manuscript.
%%%%%%%%%%%% Supplementary Methods %%%%%%%%%%%%
%\footnotesize
%\section*{Methods}

%%%%%%%%%%%%% Acknowledgements %%%%%%%%%%%%%
%\footnotesize
%\section*{Acknowledgements}

%%%%%%%%%%%%%%   Bibliography   %%%%%%%%%%%%%%
\normalsize
\bibliography{refs}

%%%%%%%%%%%%  Supplementary Figures  %%%%%%%%%%%%
%\clearpage

%%%%%%%%%%%%%%%%   End   %%%%%%%%%%%%%%%%
%\end{multicols}  % Method B for two-column formatting (doesn't play well with line numbers), comment out if using method A
\end{document}